\documentclass[12pt]{article}
\usepackage{axodraw}
\usepackage{pstricks}
\usepackage{color}

\usepackage{array}
\usepackage{epsfig}
\usepackage{amssymb}
\usepackage{graphics,graphpap}
\usepackage{amssymb}
\usepackage{amsmath}
\usepackage{slashed}
\usepackage{dsfont}

\newcommand{\tev}{\, {\rm TeV}}
\newcommand{\gev}{\, {\rm GeV}}

\newcommand{\Heff}{{\cal H}_\text{ eff}}

\newcommand{\be}{\begin{equation}}
\newcommand{\ee}{\end{equation}}
\newcommand{\bea}{\begin{eqnarray}}
\newcommand{\eea}{\end{eqnarray}}

\newcommand{\bi}{\begin{itemize}}
\newcommand{\ei}{\end{itemize}}
\newcommand{\ord}{{\cal O}}

\usepackage{graphicx}

\medskipamount 

\usepackage{fancyhdr}
\advance \headheight by 3.0truept       

\newlength{\textlength}
\newlength{\overlinelength}

\def\bbuildrel#1_#2^#3{\mathrel{\mathop{\kern 0pt#1}\limits_{#2}^{#3}}}
 \def\s#1{\setbox0=\hbox{$#1$}%
   \rlap{\ifdim\wd0>.7em\kern.22\wd0\else\kern.1\wd0\fi /}#1}


\begin{document}

\begin{titlepage}
\begin{flushright}
\begin{tabular}{l}
TUM-HEP-824/12\\
FLAVOUR(267104)-ERC-7
\end{tabular}
\end{flushright}
\vskip1.25cm
\begin{center}
{\Large \bf \boldmath
Complete NLO QCD Corrections  
for Tree Level 
$\Delta F = 2$ 
FCNC Processes:

Colourless Gauge Bosons and Scalars}
\vskip1.3cm
{\bf
Andrzej J. Buras$^{a,b}$ and 
Jennifer Girrbach$^{b,c}$}
\vskip0.5cm
$^a$ Physik Department, Technische Universit{\"a}t M{\"u}nchen,
D-85748 Garching, Germany
\\
$^b$ TUM-IAS, Lichtenbergstr. 2a, D-85748 Garching, Germany\\
$^c$ Excellence Cluster Universe, TUM, Boltzmannstra\ss{}e 2, D-85748 Garching\\
\vskip0.1cm


\vskip0.1cm

{\large\bf Abstract\\[10pt]} \parbox[t]{\textwidth}{

Anticipating the important role of tree level FCNC processes in the indirect search for new physics 
at distance scales as short as $10^{-19}-10^{-21}$~m, we present complete NLO QCD corrections to
tree level $\Delta F=2$ processes mediated by heavy colourless gauge bosons and scalars. Such contributions can be present at 
the fundamental level when the GIM mechanism is absent as in numerous 
 $Z^\prime$ models, gauged flavour models with new very heavy neutral 
gauge bosons 
and Left-Right symmetric models with 
heavy neutral scalars. 

They can also be generated at one loop in models 
having GIM at the fundamental level and Minimal Flavour Violation of which 
Two-Higgs Doublet models with and without supersymmetry are the best known 
examples. In models containing vectorial heavy fermions that mix with the standard chiral quarks 
and models in which $Z^0$ and SM neutral Higgs $H^0$ mix with new heavy 
gauge bosons and scalars in the process of electroweak symmetry breaking also 
tree-level  $Z^0$ and SM neutral Higgs $H^0$ contributions to $\Delta F=2$ 
processes are possible. In all these extensions new local operators absent in 
the SM are generated having Wilson coefficients that are generally much 
stronger affected by renormalization group QCD effects than it is the case 
of the SM operators.
The new aspect of our work is the calculation of $\mathcal{O}(\alpha_s)$
corrections to matching conditions for the Wilson coefficients of the contributing operators in the NDR-$\overline{\text{MS}}$ scheme that can be used in 
all models listed above.
This
allows to reduce certain unphysical scale and renormalization scheme
dependences in the existing NLO calculations. We show explicitly how our results can be combined
with the analytic formulae for the so-called $P_i^a$ QCD factors that include both hadronic matrix
elements of contributing operators and renormalization group evolution from high energy to low
energy scales. For the masses of heavy gauge bosons and scalars $\ord(1)\tev$ 
the remaining unphysical scale dependences for the mixing 
amplitudes $M_{12}$ 
are reduced typically from $10-25\%$, depending on the operator  considered, 
down to $1-2\%$.
}

\vfill

\end{center}
\end{titlepage}

\setcounter{footnote}{0}

\newpage

\section{Introduction}

The absence of tree level flavour changing neutral currents (FCNC) within the Standard Model (SM), known under
the name of the GIM mechanism \cite{PhysRevD.2.1285}, played a very important role in the
construction of this
model and undoubtedly contributed to its successes in an important manner. Not only are tree level FCNC
processes mediated by $Z$ boson and the Higgs absent in this model, but also the breakdown of the GIM mechanism at
the one-loop level, governed by the hierarchical structure of quark masses and of the CKM matrix, appears to be an
adequate description of existing data on FCNC processes within the experimental and theoretical uncertainties.

Beyond the SM the GIM mechanism ceases to be a general property and there exist a number of popular models in
which FCNC processes take place already at the tree level. The best known are 
various versions of the so-called $Z^\prime$ models \cite{Langacker:2008yv}
in which new neutral heavy weak boson ($Z^\prime$) 
mediate FCNC processes already at tree level. Similarly the heavy 
gauge bosons in the gauged flavour models \cite{Grinstein:2010ve} imply FCNCs at the tree-level.
This is also the case in models based
on left-right symmetry, where tree level heavy neutral Higgs exchanges 
contribute to $\Delta F=2$ amplitudes. 

Effective tree-level contributions to $\Delta F=2$ observables can  also be generated at one loop in models 
having GIM at the fundamental level and Minimal Flavour Violation of which 
Two-Higgs Doublet Models with and without supersymmetry are the best known 
examples. In models with heavy vectorial fermions that mix with the standard 
chiral quarks 
and models in which $Z^0$ and SM neutral Higgs $H^0$ mix with new heavy 
gauge bosons and scalars in the process of electroweak symmetry breaking also 
tree-level  $Z^0$ and SM neutral Higgs $H^0$ contributions to $\Delta F=2$ 
processes are possible. In all these extensions new local operators absent in 
the SM are generated having Wilson coefficients that are generally much 
stronger affected by renormalization group QCD effects than it is the case 
of the SM operators.

Extensive model independent analyses of various authors
 of $\Delta F = 2$ processes, in particular in \cite{Bona:2007vi,Antonelli:2009ws,Isidori:2010kg},
demonstrate very clearly that in the presence of $\mathcal{O}(1)$ FCNC couplings of heavy gauge bosons and
scalars, their masses must be above $100~$TeV, corresponding to distance scales as short as $10^{-21}~$m, in
order to satisfy the present experimental bounds. With couplings significantly suppressed, these
masses can be lowered to the $1~\text{TeV}-10~\text{TeV}$ range and 
even lower scales, which are in the LHC reach.

While until now no  definite signs of new physics have been observed at 
the LHC, we expect that in the coming years new phenomena, new particles 
and forces, will be discovered and their nature tested both in high energy 
processes governed by ATLAS and CMS and low-energy high precision experiments
with prominent role played by LHCb, Belle II, $K^+\to\pi^+\nu\bar\nu$ experiment at CERN and later by the 
Super-B factory in Rome and the X--project at Fermilab.

It is conceivable that these experiments will answer some of the present 
questions, simultaneously opening new ones that will require to search 
for new physics beyond the reach of the LHC. While loop diagrams, like 
penguin diagrams of various sorts and box diagrams dominated the physics of 
flavour 
changing  neutral current (FCNC) processes 
in the last thirty years both within the SM and several of its extensions, 
we should hope that  
in the case of new particles 
with masses above $10\tev$ this role will be taken over by tree-level diagrams. 
The reason is simple. Internal particles with such large masses, if hidden 
in loop diagrams, will quite generally imply very small effects that will 
be very difficult to measure. On the other hand tree diagrams could still 
provide a large window to these very short distance scales.

Anticipating this future role of tree level diagrams we make another look 
at the NLO QCD corrections to $\Delta F=2$ processes like $K^0-\bar K^0$ 
and $B^0_{d,s}-\bar B^0_{d,s}$ mixings mediated by tree level heavy neutral 
gauge bosons 
and scalars. New contributions of this type imply the presence of 
new four-fermion operators in addition to the SM $(V-A)\times(V-A)$ operator. 
They have been classified in \cite{Bagger:1997gg,Ciuchini:1997bw,Ciuchini:1998ix,Buras:2000if} and we
will list them below in
the 
basis of \cite{Buras:2000if}. 

Concerning QCD corrections, what is known are the renormalization group 
evolution matrices at the NLO level and the values of the hadronic matrix 
elements calculated using lattice methods. This 
information allows to study QCD effects in $\Delta F=2$ processes in 
a meaningful manner because only at the NLO level the Wilson coefficients can 
be properly combined with the hadronic matrix elements calculated by 
lattice methods at low energy scales.

Now as pointed out in \cite{Buras:2001ra}
instead of evaluating the hadronic matrix elements at the 
low energy scale we can choose to evaluate them at the high scale $\mu_{\rm in}$ at which heavy particles are 
integrated out. 
Thus the amplitude for a given $M-\overline{M}$ mixing ($M= K, B_d,B_s$) is given simply by
\be\label{amp6}
A(M\to \overline{M})=\langle\overline{M}|\Heff^{\Delta F=2}|M\rangle =
\kappa\sum_{i,a} C^a_i(\mu_{\rm in})\langle \overline{M} |Q^a_i(\mu_{\rm in})|M\rangle\,,
\ee
with $\kappa$ specified below. 
Here the sum runs over all the contributed operators which will be listed 
in Section~\ref{Sec2}. The 
matrix elements for $B^0_d-\bar B^0_d$ mixing are for instance given then as 
follows \cite{Buras:2001ra,Gorbahn:2009pp} 
\be\label{eq:matrix}
\langle \bar B_d^0|Q_i^a(\mu_{\rm in})|B_d^0\rangle = \frac{1}{3}m_{B_d} F_{B_d}^2 P_i^a(B_d)\equiv \bar P^a_i(\mu_{\rm in})\,,
\ee
where the coefficients $P_i^a(B_d)$, in which the scale $\mu_{\rm in}$ has been 
suppressed,
collect compactly all RG effects from scales below $\mu_{\rm in}$ as well as
hadronic matrix elements obtained by lattice methods at low energy scales.
Analytic formulae for these coefficients are given in \cite{Buras:2001ra} 
while the recent application of this method can be found in 
\cite{Buras:2010mh,Buras:2010zm,Buras:2010pz,Blanke:2011ry,Buras:2011wi}. As the 
Wilson coefficients $ C_i(\mu_{\rm in})$ depend directly on the loop functions, 
tree diagram results
and fundamental parameters of a given theory, this formulation is very 
transparent and interesting short distance NP effects are not hidden 
by complicated QCD effects. 

In this approach the hadronic matrix elements in   Eq.~(\ref{amp6}) are usually calculated in a particular
renormalization scheme at the matching scale $\mu_\text{in}$. This scale, while being 
of the order 
of the masses of heavy particles that have been integrated out,  does not 
have to be equal to these masses. As the amplitude on the l.h.s of  Eq.~(\ref{amp6})
cannot depend on the choice of the renormalization scheme and on  the precise value of the scale 
$\mu_\text{in}$, these unphysical dependences have to be cancelled by the ones 
present also in the Wilson 
coefficients $C_i^a(\mu_\text{in})$. To this end these coefficients have to be 
known at the NLO level which requires the calculation of $\ord(\alpha_s)$ 
corrections to penguin diagrams, box diagrams and in particular tree diagrams 
in the full theory and matching this calculation to the corresponding 
effective theory.

Now in most applications to date the coefficients  $C_i^a(\mu_\text{in})$ 
in the extensions of the SM have 
been calculated at the leading order, leaving some left-over unphysical 
scheme and scale dependences in the resulting physical amplitudes. While 
presently these dependences are significantly smaller than the present
 uncertainties 
in the evaluation of the hadronic matrix elements, the situation could change in this decade. In the SM these coefficients are known at the NLO and  in a few 
processes at the NNLO level. An up-to-date review can be found in 
\cite{Buras:2011we}.

The goal of our paper is the evaluation of $C_i^a(\mu_\text{in})$ for the cases
of  tree-level colourless neutral gauge boson and neutral scalar exchanges at the 
NLO level. This amounts to the calculation of $\ord(\alpha_s)$ corrections to 
the tree diagrams in question. We will also show explicitly how our results should be 
combined  with the coefficients $P_i^a$ so that our final results will 
be mixing amplitudes at the NLO level that are general enough to be used 
for any model in which tree-level contributions to $\Delta F=2$ processes 
from colourless neutral gauge boson and 
scalar exchanges are present. The analysis of coloured gauge bosons and 
scalars is in progress but the calculations in  this case are 
more involved and we will present them  
in a separate publication \cite{BG2}.

Our paper is organized as follows: In Section~\ref{Sec2} we recall the general structure 
of the effective Hamiltonian for $\Delta F=2$ processes and we give the full 
list of four-fermion operators that contribute to these transitions. In 
Section~\ref{Sec3}, the most important section of our paper, we describe the calculation 
of one-loop QCD corrections to the coefficients $C_i^a$ in the 
NDR-$\overline{\text{MS}}$ scheme
and collect our results. The separate results for the amplitudes in 
the full theory and the effective theory given in the Appendices C and D should 
enable interested readers to check our calculations. In Section~\ref{Sec4} 
we demonstrate analytically that the new contributions remove unphysical scale dependences in the  formulae present in the literature.
In Section~\ref{Sec5} we combine our results with the $P_i^a$ QCD factors of
\cite{Buras:2001ra}
obtaining in this manner the complete NLO results for the $\Delta F=2$ 
amplitudes governed by tree-level exchanges of colourless 
gauge bosons and scalars. 
This section can be considered as a compendium of the {\it master formulae} for
tree-level contributions to  
the mixing amplitudes at the NLO level 
that are valid in {\it any} extension of the SM in which 
such contributions are present.
In Section~\ref{Sec6}
we demonstrate numerically that the resulting amplitudes practically 
do not depend on the detail 
choice of the matching scale. We conclude with a brief summary in Section~\ref{Sec7}.

 \section{Theoretical Framework}\label{Sec2}
 \subsection{Preliminaries}
While in the SM only one operator contributes to each $\Delta F=2$ transition 
in the $K$ and $B$ systems, 
in the tree level FCNC transition considered here there are eight such operators of dimension six. 
Consequently the renormalization group (RG) QCD analysis becomes more 
involved and due to the presence of right-handed and scalar currents and 
the resulting
structure of the new operators
QCD corrections play a much more  important role in 
new physics contributions than in the SM contributions. Therefore also the
unphysical renormalization scheme and scale dependences are much more 
pronounced when not all NLO QCD corrections are taken into account. 

In what follows, after listing all contributing operators we will 
summarize the effective Hamiltonian for $\Delta F=2$ transitions.

\subsection{Local Operators}
The contributing four-fermion operators 
can be split into five separate sectors, according to the chirality
of the quark fields they contain. 
For definiteness, we shall consider  operators responsible for the
$K^0$--$\bar{K}^0$ mixing. 
The operators belonging to the first
three sectors (VLL, LR and SLL) read \cite{Buras:2000if} :
\begin{align} \begin{split}
Q_1^{\rm VLL}(K) &=(\bar{s}^{\alpha} \gamma_{\mu}    P_L d^{\alpha})
(\bar{s}^{ \beta} \gamma^{\mu}    P_L d^{ \beta})\,,
\\[4mm] 
Q_1^{\rm LR}(K) &=  (\bar{s}^{\alpha} \gamma_{\mu}    P_L d^{\alpha})
(\bar{s}^{ \beta} \gamma^{\mu}    P_R d^{ \beta})\,,
\\
Q_2^{\rm LR}(K) &=  (\bar{s}^{\alpha}                 P_L d^{\alpha})
(\bar{s}^{ \beta}                 P_R d^{ \beta})\,,
\\[4mm]
Q_1^{\rm SLL}(K) &= (\bar{s}^{\alpha}                 P_L d^{\alpha})
(\bar{s}^{ \beta}                 P_L d^{ \beta})\,,
\\
Q_2^{\rm SLL}(K) &=(\bar{s}^{\alpha} \sigma_{\mu\nu} P_L d^{\alpha})
(\bar{s}^{ \beta} \sigma^{\mu\nu} P_L d^{ \beta})\,\end{split}
\label{normalK}
\end{align}
where $\sigma_{\mu\nu} = \frac{1}{2} [\gamma_{\mu}, \gamma_{\nu}]$ and
$P_{L,R} = \frac{1}{2} (1\mp \gamma_5)$. The operators belonging to the
two remaining sectors (VRR and SRR) are obtained from $Q_1^{\rm VLL}$ and
$Q_i^{\rm SLL}$ by interchanging $P_L$ and $P_R$. In the SM only the 
operator $Q_1^{\rm VLL}(K)$ is present.
The operators relevant for  $B_q$ ($q = d,s$) are obtained by replacing 
in (\ref{normalK}) $K$ by $B_q$,
 $s$ by $b$ and
$d$ by $q$.

\subsection{Effective Hamiltonian}
The effective Hamiltonian for $\Delta F=2$ transitions can be written 
in a general form as follows 
\be\label{Heff-general}
\Heff^{\Delta F=2} =\kappa
\sum_{i,a}C_i^a(\mu)Q_i^a\,,
\ee
where $Q_i^a$ are the operators given in Eq.~(\ref{normalK}) and
$C_i^a(\mu)$ their Wilson coefficients evaluated at a scale $\mu$ at which 
the hadronic matrix elements are evaluated. The overall factor $\kappa$ depends on the contributing particles
 and will be chosen such that for non-vanishing Wilson coefficients 
$C_i^a(\mu_{\rm in}) = 1$ in the LO. The scale $\mu$ can be the low 
energy scale $\mu_L$ at which actual lattice calculations 
are performed or any other 
scale, in particular the matching scale $\mu_\text{in}$ as in Eq.~(\ref{amp6}). 
In this case the matrix elements are obtained by evolving by means of
RG equations the lattice results from $\mu_L$ to $\mu_\text{in}$. The result 
of this evolution is given in Eq.~(\ref{eq:matrix}). The general NLO formulae 
for the coefficients $P_i^a$ as functions of the QCD coupling 
constant and the hadronic $B_i$ parameters calculated by lattice methods are
presented in  \cite{Buras:2001ra}. 

As already emphasized in the Introduction, this formulation is 
very powerful as it applies to any extension of the SM. What distinguishes 
various NP scenarios are
\begin{itemize}
\item 
the contributing operators $Q_i^a$,
\item
their 
Wilson coefficients $ C_i^a(\mu_{\rm in})$ which depend directly on 
the fundamental parameters of a given theory.  
\end{itemize}

The results of 
tree-level and loop calculations performed at the matching scale $\mu_\text{in}$ 
at which the heavy particles are integrated out depend explicitly 
on the these fundamental parameters which allows to see very transparently 
the  short distance NP effects that are not hidden 
by complicated QCD effects which necessarily take place between high energy 
and low energy scales. 

\subsection{The Operator Structure from Tree Level Exchanges}
In the present paper we will consider FCNC $\Delta F=2$ amplitudes 
generated through tree-level very heavy gauge boson and scalar exchanges 
independently whether flavour violating neutral couplings in question 
have been  generated at 
the fundamental level or through loop corrections. 
The only assumption that we will make in the present paper is that exchanged 
neutral gauge bosons and scalars are colourless as in many NP scenarios 
listed above.

It is instructive to compare the operator structures in the effective 
Hamiltonian for $\Delta F=2$ transitions at scales $\ord(\mu_{\rm in})$ 
resulting from tree-level 
exchanges that differ when gauge bosons and scalars with colour or 
without colour are exchanged. We have:
\begin{itemize}
\item
A tree level exchange of a colourless gauge boson with LH and RH couplings 
generates at $\mu_\text{in}$ the operators $Q_1^{\rm VLL}$, $Q_1^{\rm VRR}$ and $Q_1^{\rm LR}$. This is an 
example of $Z^\prime$ models and  gauge flavour models \cite{Grinstein:2010ve}.
Also tree-level $Z^0$ can be generated in some extensions of the SM, in 
particular when new heavy neutral gauge bosons mix with $Z^0$ and heavy vectorial fermions mix with chiral quarks of the SM.
When QCD corrections at the $\mu_\text{in}$ are taken into account also the 
operator $Q_2^{\rm LR}$ is generated but its Wilson coefficient is suppressed 
by $\alpha_s(\mu_\text{in})$ relative to other operators as we will see in the next 
section. 
\item
A tree level exchange of  a gauge boson carrying colour generates the operators
$Q_1^{\rm VLL}$, $Q_1^{\rm VRR}$, $Q_1^{\rm LR}$ and $Q_2^{\rm LR}$ even 
without including QCD corrections. An example is the 
tree-level exchange of the KK-gluon in the RS models.
\item
A tree level exchange of a colourless Higgs scalar generates the 
operators  $Q_1^{\rm SLL}$, $Q_1^{\rm SRR}$ and $Q_2^{\rm LR}$. 
When QCD corrections at the $\mu_\text{in}$ are taken into account also the 
operators $Q_2^{\rm SLL}$, $Q_2^{\rm SRR}$ and $Q_1^{\rm LR}$
are generated but their Wilson coefficient are suppressed 
by $\alpha_s(\mu_\text{in})$ relative to other operators as we will see in the next 
section. Such heavy scalars are present in supersymmetric models and 
in left-right symmetric models. Tree level exchanges of the SM $H^0$ could 
also be generated in certain models.
\item
A tree level exchange of a Higgs scalar carrying colour generates the operators
$Q_{1,2}^{\rm SLL}$, $Q_{1,2}^{\rm SRR}$ and $Q_{1,2}^{\rm LR}$ even without 
the inclusion of QCD corrections.
\end{itemize}

As already stated before we concentrate here on the colourless gauge bosons 
and scalars. The case of coloured gauge bosons and scalars will be discussed 
elsewhere \cite{BG2}.

\section{Matching Conditions}\label{Sec3}
\subsection{Preliminaries}

The calculations of $\ord(\alpha_s)$ QCD corrections to Wilson coefficients 
are by now standard and have been described in several papers. 
In particular in Section 5.4.2 of \cite{Buras:1998raa} all necessary steps have been 
presented in detail in the case of charged currents within the SM, 
while \cite{Buras:1990fn} presents the calculation of  $\ord(\alpha_s)$ corrections 
to $C_1^\text{VLL}$ within the SM. The novel feature of the calculations 
present below when compared with these two papers is the appearance of 
new operators but the procedure is the same:

{\bf Step 1}

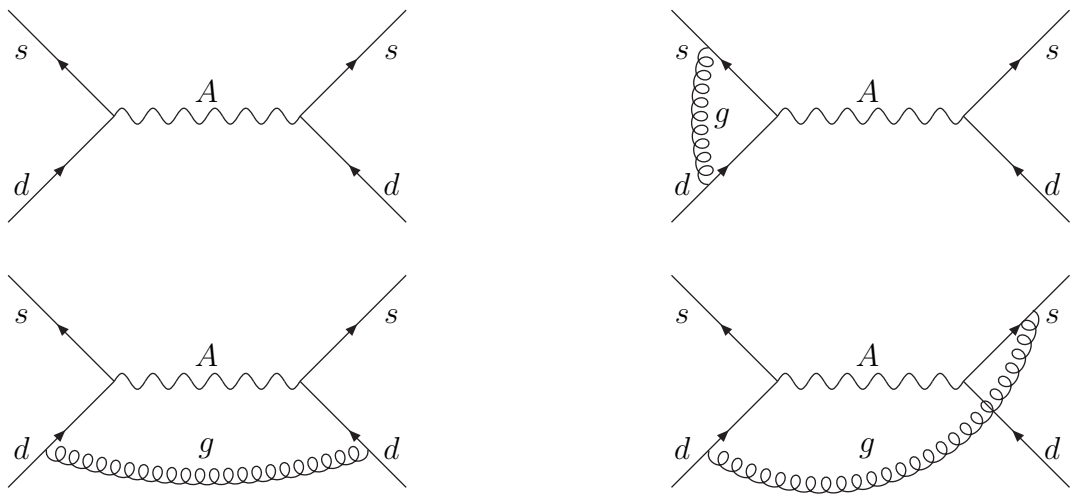
\begin{figure}[!tbh]
\begin{center}
\scalebox{1.0}{
    \begin{picture}(400,200)(0,0)
      \SetColor{Black}

	\ArrowLine(40,150)(0,190)\Text(5,175)[]{$ s$}
	\ArrowLine(0,110)(40,150)\Text(5,125)[]{$d$}
	\Photon(40,150)(110,150){3}{6}\Text(75,160)[c]{$A$}
	\ArrowLine(110,150)(150,190)\Text(145,175)[]{$ s$}
	\ArrowLine(150,110)(110,150)\Text(145,125)[]{$  d$}
	
	\ArrowLine(290,150)(250,190)\Text(255,175)[]{$ s$}
	\ArrowLine(250,110)(290,150)\Text(255,125)[]{$d$}
	\Photon(290,150)(360,150){3}{6}\Text(325,160)[c]{$A$}
	\ArrowLine(360,150)(400,190)\Text(395,175)[]{$ s$}
	\ArrowLine(400,110)(360,150)\Text(395,125)[]{$  d$}
	\GlueArc(360,150)(99.45,165,195){3}{9}\Text(270,150)[]{$g$}

	\ArrowLine(40,50)(0,90)\Text(5,75)[]{$ s$}
	\ArrowLine(0,10)(40,50)\Text(5,25)[]{$d$}
	\Photon(40,50)(110,50){3}{6}\Text(75,60)[c]{$A$}
	\ArrowLine(110,50)(150,90)\Text(145,75)[]{$ s$}
	\ArrowLine(150,10)(110,50)\Text(145,25)[]{$  d$}
	\GlueArc(75,200)(186,-109,-71){3}{22}\Text(75,25)[c]{$g$}

	\ArrowLine(290,50)(250,90)\Text(255,75)[]{$s$}
	\ArrowLine(250,10)(290,50)\Text(255,25)[]{$d$}
	\Photon(290,50)(360,50){3}{6}\Text(325,60)[c]{$A$}
	\ArrowLine(360,50)(400,90)\Text(395,75)[]{$ s$}
	\ArrowLine(400,10)(360,50)\Text(395,25)[]{$  d$}
	\GlueArc(307.6,90.6)(79.75,-123,-10){3}{26}\Text(325,25)[c]{$g$}

         \end{picture}}
\caption{\it Tree level diagram and one-loop QCD corrections to $\Delta S = 2$ transition mediated by a gauge
boson in the full theory. The mirror diagrams are not shown.}\label{fig:fullgauge}
\end{center}
\end{figure}

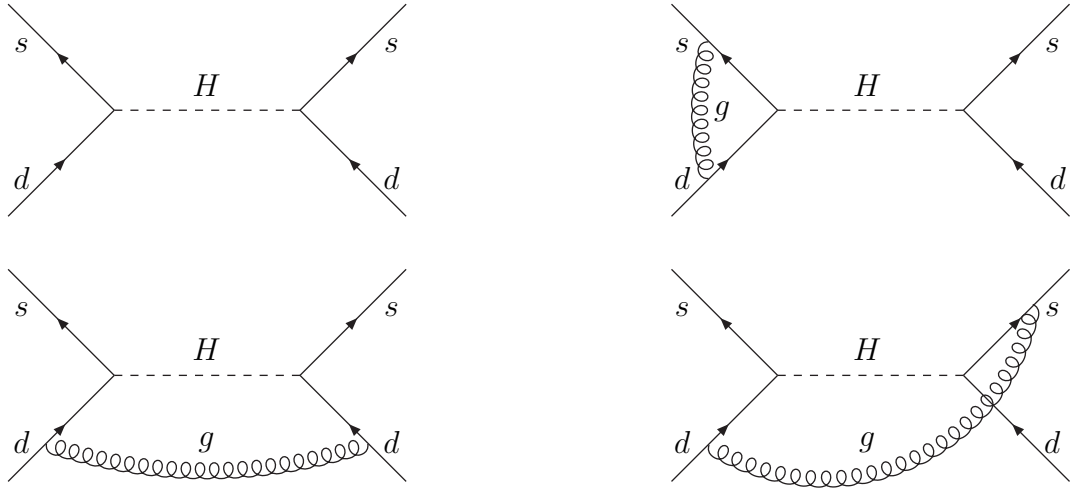
\begin{figure}[!tbh]
\begin{center}
\scalebox{1.0}{
    \begin{picture}(400,200)(0,0)
      \SetColor{Black}

	\ArrowLine(40,150)(0,190)\Text(5,175)[]{$s$}
	\ArrowLine(0,110)(40,150)\Text(5,125)[]{$d$}
	\DashLine(40,150)(110,150){3}\Text(75,160)[c]{$H$}
	\ArrowLine(110,150)(150,190)\Text(145,175)[]{$ s$}
	\ArrowLine(150,110)(110,150)\Text(145,125)[]{$  d$}

	\ArrowLine(290,150)(250,190)\Text(255,175)[]{$ s$}
	\ArrowLine(250,110)(290,150)\Text(255,125)[]{$d$}
	\DashLine(290,150)(360,150){3}\Text(325,160)[c]{$H$}
	\ArrowLine(360,150)(400,190)\Text(395,175)[]{$ s$}
	\ArrowLine(400,110)(360,150)\Text(395,125)[]{$ \ d$}
	\GlueArc(360,150)(99.45,165,195){3}{9}\Text(270,150)[]{$g$}

	\ArrowLine(40,50)(0,90)\Text(5,75)[]{$ s$}
	\ArrowLine(0,10)(40,50)\Text(5,25)[]{$d$}
	\DashLine(40,50)(110,50){3}\Text(75,60)[c]{$H$}
	\ArrowLine(110,50)(150,90)\Text(145,75)[]{$ s$}
	\ArrowLine(150,10)(110,50)\Text(145,25)[]{$  d$}
	\GlueArc(75,200)(186,-109,-71){3}{22}\Text(75,25)[c]{$g$}

	\ArrowLine(290,50)(250,90)\Text(255,75)[]{$ s$}
	\ArrowLine(250,10)(290,50)\Text(255,25)[]{$d$}
	\DashLine(290,50)(360,50){3}\Text(325,60)[c]{$H$}
	\ArrowLine(360,50)(400,90)\Text(395,75)[]{$ s$}
	\ArrowLine(400,10)(360,50)\Text(395,25)[]{$  d$}
	\GlueArc(307.6,90.6)(79.75,-123,-10){3}{26}\Text(325,25)[c]{$g$}

         \end{picture}}
\caption{\it Tree level diagram and one-loop QCD corrections to $\Delta F = 2$ transition mediated by a scalar
particle in the full theory. The mirror diagrams are not shown.}\label{fig:fullscalar}
\end{center}
\end{figure}

We first calculated the amplitudes in the full theory. They are given in the 
case of a gauge boson exchange and a scalar exchange in Figs.~\ref{fig:fullgauge} and \ref{fig:fullscalar},
respectively.
In the presence of massless gluons one encounters infrared divergences. 
We have regulated these 
divergences by a common external momentum $p$ with $-p^2>0$ for all external 
massless fields as done in \cite{Buras:1998raa}. Equally well they could be
regulated by setting all external momenta 
to zero but giving the external quarks non-vanishing masses as done in \cite{Buras:1990fn}. As the Wilson coefficients cannot depend on the employed 
infrared regulator, 
the same result should be obtained in both cases. In fact in the case of 
gauge boson exchanges we have performed also calculations with the mass 
regulator obtaining the same results for the Wilson coefficients.

The ultraviolet 
divergences present in the vertex diagrams in 
 Figs.~\ref{fig:fullgauge} and \ref{fig:fullscalar}
have been regulated 
using dimensional regularization with anti-commuting $\gamma_5$ in $4-2\epsilon$
dimensions.

{\bf Step 2}

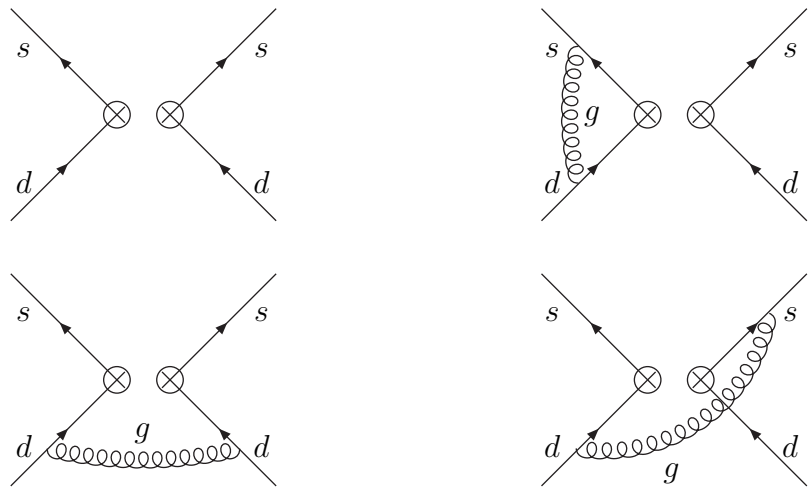
\begin{figure}[!tbh]
\begin{center}
\scalebox{1.0}{
    \begin{picture}(400,200)(0,0)
      \SetColor{Black}

	\ArrowLine(90,150)(50,190)\Text(55,175)[]{$ s$}
	\ArrowLine(50,110)(90,150)\Text(55,125)[]{$d$}
	\CArc(90,150)(5,0,360)  
	\Line(90,150)(93,153)\Line(90,150)(93,147)

	\ArrowLine(110,150)(150,190)\Text(145,175)[]{$ s$}
	\ArrowLine(150,110)(110,150)\Text(145,125)[]{$  d$}
	\CArc(110,150)(5,0,360) 
	\Line(110,150)(107,153)\Line(110,150)(107,147)

	\ArrowLine(290,150)(250,190)\Text(255,175)[]{$ s$}
	\ArrowLine(250,110)(290,150)\Text(255,125)[]{$d$}
	\CArc(290,150)(5,0,360) 
	 \Line(290,150)(293,153)\Line(290,150)(293,147)
\GlueArc(360,150)(99.45,165,195){3}{9}\Text(270,150)[]{$g$}
	\ArrowLine(310,150)(350,190)\Text(345,175)[]{$ s$}
	\ArrowLine(350,110)(310,150)\Text(345,125)[]{$  d$}
	\CArc(310,150)(5,0,360) 
	\Line(310,150)(307,153)\Line(310,150)(307,147)

	\ArrowLine(90,50)(50,90)\Text(55,75)[]{$ s$}
	\ArrowLine(50,10)(90,50)\Text(55,25)[]{$d$}
	\CArc(90,50)(5,0,360)  
	\Line(90,50)(93,53)\Line(90,50)(93,47)
      \GlueArc(100,175)(155.23,-103.5,-76.5){3}{13}\Text(100,30)[]{$g$}
	\ArrowLine(110,50)(150,90)\Text(145,75)[]{$ s$}
	\ArrowLine(150,10)(110,50)\Text(145,25)[]{$  d$}
	\CArc(110,50)(5,0,360) 
	\Line(110,50)(107,53)\Line(110,50)(107,47)

	\ArrowLine(290,50)(250,90)\Text(255,75)[]{$ s$}
	\ArrowLine(250,10)(290,50)\Text(255,25)[]{$d$}
	\CArc(290,50)(5,0,360) 
	 \Line(290,50)(293,53)\Line(290,50)(293,47)
 \GlueArc(274,86)(62.8,-100,-10){3}{16}\Text(300,15)[]{$g$}
	\ArrowLine(310,50)(350,90)\Text(345,75)[]{$ s$}
	\ArrowLine(350,10)(310,50)\Text(345,25)[]{$  d$}
	\CArc(310,50)(5,0,360) 
	\Line(310,50)(307,53)\Line(310,50)(307,47)

         \end{picture}}
\caption{\it Leading order and one-loop diagrams in the effective theory. 
The mirror diagrams are not shown.}\label{fig:eff}
\end{center}
\end{figure}

We have calculated the matrix elements of contributing operators by 
evaluating the diagrams in Fig.~\ref{fig:eff} making the same
assumptions about the 
external fields as in the first step. In contrast to step 1 one 
has to renormalize the operators. This we do in the $\overline{\text{MS}}$ 
renormalization scheme with anti-commuting $\gamma_5$, which corresponds 
to the NDR scheme of \cite{Buras:1989xd} used also in \cite{Buras:2000if} and \cite{Buras:2001ra}.

{\bf Step 3}

 We finally inserted the results of the two steps above into the formula 
 like the one in Eq.~(\ref{amp6}) and comparing the coefficients of 
 operators appearing on the l.h.s (full theory) and r.h.s (effective 
 theory) we found the coefficients $C_i^a(\mu_{\rm in})$. As these 
 coefficients cannot depend on the infrared behaviour of the theory, 
 the dependences on $p^2$  found in the first two 
 steps have to cancel each other in the evaluation of   $C_i^a(\mu_{\rm in})$. 
 Indeed we verified this explicitly. The interested reader can do this 
as well by inspecting our intermediate results that we present in Appendices 
C and D.  
The appearance of the renormalization 
 scale $\mu_{\rm in}$ can be traced to the use of dimensional regularization 
 and the renormalization in the $\overline{\text{MS}}$ scheme.

 Very often in analyses in which NP contributions are governed by 
 box diagrams the overall factor in front of the sum in Eq.~(\ref{amp6}) is 
 chosen as in the SM. However, in our analysis it will be more convenient 
to use 
 in each case the normalization in which the Wilson coefficient of the
leading operator evaluated at the matching scale is 
 equal to unity in the absence of QCD corrections. In this manner the 
applications of our formulae in various models will be facilitated. In what 
 follows we will first present the general structure of the effective Hamiltonian  in each case. Subsequently we will list our results for the 
 Wilson coefficients including $\ord(\alpha_s)$ corrections.

\subsection{Results}

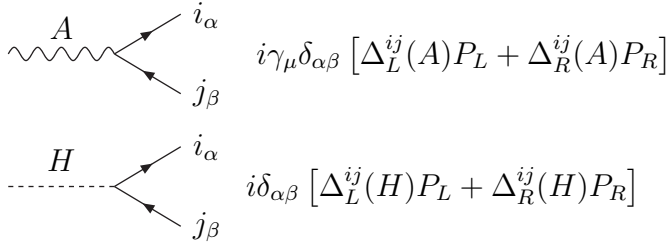
\begin{figure}[!htb]
\scalebox{1.0}{
    \begin{picture}(100,150)(0,-50)
      \SetColor{Black}

	\Photon(0,70)(40,70){2.5}{4.5}\ArrowLine(40,70)(65,85)\ArrowLine(65,55)(40,70) 
	\Text(20,80)[c]{$A$}\Text(70,85)[l]{$i_\alpha$}\Text(70,55)[l]{$j_\beta$}\Text(90,70)[l]{
$i\gamma_\mu\delta_{\alpha\beta}\left[
\Delta_L^{ij}(A)P_L+\Delta_R^{ij}(A)P_R\right]$}

	\DashLine(0,20)(40,20){2}\ArrowLine(40,20)(65,35)\ArrowLine(65,5)(40,20)  
	\Text(20,30)[c]{$H$}\Text(70,35)[l]{$i_\alpha$}\Text(70,5)[l]{$j_\beta$}\Text(90,20)[l]{$i\delta_{
\alpha\beta}\left[ \Delta_L^{ ij}
(H)P_L+\Delta_R^{ij}(H)P_R\right]$}
        \end{picture}}
\vskip-1.5cm
\caption{\it Feynman rules for colourless neutral gauge boson $A$ with mass $M_A$, 
and neutral colourless scalar particle $H$
with mass $M_H$, where $i,\,j$ denote different
quark flavours and $\alpha,\,\beta$ the colours.}\label{fig:Feynman}
\end{figure}

Defining the general couplings through 
the  Feynman rules in Fig.~\ref{fig:Feynman} we find the following results.

\subsubsection{Colourless gauge boson}\label{CLGB}

\begin{align}\begin{split}
 \mathcal{H}_\text{eff}^{\Delta S = 2}  =&
\frac{(\Delta_L^{sd}(A))^2}{2M_A^2}C_1^\text{VLL}(\mu)Q_1^\text{VLL}+\frac{(\Delta_R^{sd}(A))^2}{2M_A^2}
C_1^\text{VRR}(\mu)Q_1^\text{VRR} \\
&+\frac{\Delta_L^{sd}(A)\Delta_R^{sd}(A)}{
M_A^2} \left [ C_1^\text{LR}(\mu) Q_1^\text{LR} +C_2^\text{LR}(\mu) Q_2^\text{LR} \right]\,.\end{split}
\end{align}
The operator $ Q_2^\text{LR}$ is only generated at one-loop but not present at tree level. Consequently its
Wilson coefficient is $\mathcal{O}(\alpha_s)$. We find for an arbitrary 
number of colours $N$
{\allowdisplaybreaks
\begin{align}
\begin{split}
C_1^\text{VLL}(\mu)=C_1^\text{VRR}(\mu) & =
1+\frac{\alpha_s}{4\pi}\left(1-\frac{1}{N}\right)\left(-3\log\frac{M_A^2}{\mu^2}+\frac{11}{2}\right)\\
& = 1+\frac{\alpha_s}{4\pi}\left(-2\log\frac{M_A^2}{\mu^2}+\frac{11}{3}\right)\,,\end{split}\\
\begin{split}
 C_1^\text{LR}(\mu) & =
1+\frac{\alpha_s}{4\pi}\frac{3}{N}\left(-\log\frac{M_A^2}{\mu^2}-\frac{1}{6}\right)\\
& =1+\frac{\alpha_s}{4\pi}
\left(-\log\frac{M_A^2}{\mu^2}-\frac{1}{6}\right)\,,\end{split}\\
C_2^\text{LR}(\mu) &=\frac{\alpha_s}{4\pi}\left(-6\log\frac{M_A^2}{\mu^2}-1\right)\,.
\end{align}}%

\subsubsection{Colourless scalar}\label{CLS}

\begin{align}\begin{split}
 \mathcal{H}_\text{eff}^{\Delta S = 2} =&
-\frac{(\Delta_L^{sd}(H))^2}{2M_H^2}\left[C_1^\text{SLL}(\mu)Q_1^\text{SLL} +C_2^\text{SLL}(\mu)Q_2^\text{
SLL}  \right]\\
&-\frac{(\Delta_R^{sd} (H))^2 } {2 M_H^2 }
\left[C_1^\text{SRR}(\mu)Q_1^\text{SRR} +C_2^\text{SRR}(\mu)Q_2^\text{SRR} \right]\\
&-\frac{\Delta_L^{sd}(H)\Delta_R^{sd}(H)}{
M_H^2} \left [ C_1^\text{LR}(\mu) Q_1^\text{LR} +C_2^\text{LR}(\mu) Q_2^\text{LR} \right]\,.\end{split}
\end{align}
The operators $Q_2^\text{SLL/SRR} $ and $Q_1^\text{LR}$ are generated by QCD corrections. We find for an arbitrary number of colours $N$
{\allowdisplaybreaks
\begin{align}
C_1^\text{SLL}(\mu)= C_1^\text{SRR}(\mu)&=
1+\frac{\alpha_s}{4\pi}\left(-3\log\frac{M_H^2}{\mu^2}+\frac{9}{2}\right)\,,\\
\begin{split}
C_2^\text{SLL}(\mu) =
C_2^\text{SRR}(\mu) &=\frac{\alpha_s}{4\pi}\frac{2-N}{4N}\left(\log\frac{M_H^2}{\mu^2}-\frac{3}{2}
\right)\\
&=\frac{\alpha_s}{4\pi}\left(-\frac{1}{12}\log\frac{M_H^2}{\mu^2}+\frac{1}{8}
\right)\,,\end{split}\\
 C_1^\text{LR}(\mu)& =-\frac{3}{2}\frac{\alpha_s}{4\pi}\,,\\
C_2^\text{LR}(\mu) &=  1-\frac{\alpha_s}{4\pi}\frac{3}{N}=  1-\frac{\alpha_s}{4\pi}\,.
\end{align}}%

 The formulae presented in this subsection  are the main results of our paper.

\boldmath
\section{Renormalization Scale Dependence}\label{Sec4}
\unboldmath

One of the main virtues of our calculation of $\mathcal{O}(\alpha_s)$ corrections to Wilson coefficients at the high energy matching scale
$\mu_{\rm in}$ is the cancellation of the $\mu_{\rm in}$ dependence of the renormalization group evolution matrix by the $\mu_{\rm in}$
dependence of the Wilson
coefficients in question. This cancellation requires particular values of the coefficients of the $\log(M^2/\mu_{\rm in}^2)$ in
$C_i(\mu_{\rm in})$ where $M$ stands for the mass of a heavy gauge boson or heavy scalar involved. As this cancellation constitutes an
important test of our results
it is useful to derive a general condition on the coefficients of 
$\log(M^2/\mu_{\rm in}^2)$ in $C_i(\mu_{\rm in})$.

To this end let us look as an example at the evolution matrix 
$\hat{U}(\mu_b,\,\mu_{\rm in})$ defined by 
\be
\vec{C}(\mu_b)=\hat{U}(\mu_b,\,\mu_{\rm in})\vec{C}(\mu_{\rm in}),
\ee
Here $\vec{C}(\mu)$ is a column vector of Wilson coefficients.
Expanding then this matrix
 around the two fixed scales $m_b$ and $M$ keeping
only the logarithmic terms one obtains
\begin{align}\label{equ:hatU}
 \hat U(\mu_b,\,\mu_{\rm in}) =
\left(\mathds{1}+\frac{\alpha_s(\mu_b)}{4\pi}\frac{\hat\gamma^{(0)\top}}{2}\log\frac{\mu_b^2}{m_b^2}\right)\hat{U}(m_b,\,M)
\left(\mathds{1}+\frac{\alpha_s(\mu_{\rm in})}{4\pi}\frac{\hat\gamma^{(0)\top}}{2}\log\frac{M^2}{\mu_{\rm in}^2}\right)\,,
\end{align}
where $\hat\gamma^{(0)}$ is the coefficient of $\alpha_s$ in the one loop anomalous dimension matrix that describes the mixing of operators:
\begin{align}
 \hat\gamma = \frac{\alpha_s}{4\pi}\hat\gamma^{(0)}+\mathcal{O}(\alpha_s^2)\,.
\end{align}
Note that it is $\hat\gamma^{(0)\top}$ and not $\hat\gamma^{(0)}$ that enters (\ref{equ:hatU}).
Moreover, in the study of the $\mu_{\rm in}$ dependence in the case of the 
scalar exchange one has to take into account that in this case 
the $m^2(\mu_{\rm in})$ dependence is hidden in
the coefficients $(\Delta_{L/R}^{ij}(H))^2$.

Considering then the cases of colourless gauge bosons and scalars we find that the following quantities should be $\mu_{\rm
in}$-independent:
\begin{subequations}\label{equ:R}
\begin{align}
 R^\text{gauge} & = \hat U(\mu_b,\,\mu_{\rm in})\vec{C}(\mu_{\rm in})\,,\\
 R^\text{scalar} & = \hat U(\mu_b,\,\mu_{\rm in})\vec{C}(\mu_{\rm in})m^2(\mu_{\rm in})\,.
 \end{align}
\end{subequations}
For VLL (VRR) the column vector  $\vec{C}(\mu)$
is just a one-dimensional one, while it is
two-dimensional for LR and SLL (SRR) systems. 
We write next
\begin{align}
 \vec{C}(\mu_{\rm in}) = \vec{C}_0 - \frac{\alpha_s(\mu_{\rm in})}{4\pi}\vec{K} \log\frac{M^2}{\mu_{\rm in}^2}\,,
\end{align}
where we suppressed $\mu_{\rm in}$ independent $\mathcal{O}(\alpha_s)$ terms.
 Moreover, the leading logarithm at $\ord(\alpha_s)$ in $ m^2(\mu_{\rm in})$ 
is given in
\begin{align}
 m^2(\mu_{\rm in}) = m^2(M)\left(1+ \frac{\alpha_s(\mu_\text{in})}{4\pi}\gamma^{(0)}_m \log\frac{M^2}{\mu_{\rm in}^2}\right)\,,
\end{align}
with $\gamma^{(0)}_m$ governing the scale dependence of the quark masses in QCD.

Imposing (\ref{equ:R}), the conditions for $\vec{K}$ to ensure 
$\mu_{\rm in}$ independence of resulting amplitudes in these two cases read
\begin{subequations}\label{equ:vecK}
\begin{align}
 \vec{K}^\text{gauge}& = \frac{\hat\gamma^{(0)\top}}{2}\vec{C}_0\,,\\
\vec{K}^\text{scalar} & = \left[\frac{\hat\gamma^{(0)\top}}{2}+\gamma^{(0)}_m\mathds{1}\right]\vec{C}_0\,,
\end{align}
\end{subequations}
where $\mathds{1}$ is a unit matrix.
Thus the coefficients of logarithms in $\vec{C}(\mu_{\rm in})$ can be found without the calculation of the loop diagrams in
Fig.~\ref{fig:fullgauge}-\ref{fig:eff} but the formulae in Eq.~(\ref{equ:vecK})  serve as a useful check of our results for logarithmic
terms. These
terms are renormalization scheme independent and while cancelling 
the $\mu_{\rm in}$ dependence of $\hat U(\mu_b,\,\mu_{\rm in})$ in perturbation theory
cannot remove its renormalization scheme dependence at the NLO level. To this end the $\mathcal{O}(\alpha_s)$ non-logarithmic terms have to
be calculated which constitutes the main new result of our paper. 

In order to be able to use the relations in Eq.~(\ref{equ:vecK})  we recall the relevant one-loop anomalous dimension matrices
\cite{Buras:2000if}:
\begin{align}
 \gamma^{(0)\text{VLL}} = 6-\frac{6}{N} = 4\,,\qquad \gamma^{(0)}_m=6 C_F=3N-\frac{3}{N}=8,
\end{align}
\begin{align}\label{LRQ}
\hat{\gamma}^{(0)\text{LR}} = \begin{pmatrix}
                               \frac{6}{N} & 12\\
				0 & -6N+\frac{6}{N}     
                               \end{pmatrix} 
                           =      \begin{pmatrix}
                               \frac{6}{N} & 12\\
				0 & -2 \gamma^{(0)}_m    
                               \end{pmatrix} 
                           = \begin{pmatrix}
                               2 & 12\\
				0 & -16
                               \end{pmatrix}
\end{align}

\begin{align}\label{SLLQ}
\hat{\gamma}^{(0)\text{SLL}} = \begin{pmatrix}
                              -6 N + 6 +\frac{6}{N} & \frac{1}{2}-\frac{1}{N}\\
				-24 -\frac{48}{N} & 2 N + 6 -\frac{2}{N}     
                               \end{pmatrix} 
                                =\begin{pmatrix}
                               -2\gamma^{(0)}_m + 6  & \frac{1}{2}-\frac{1}{N}\\
				-24 -\frac{48}{N} & \gamma^{(0)}_m\frac{2}{3} + 6      
                               \end{pmatrix} 
                               = \begin{pmatrix}
                                -10 & \frac{1}{6}\\
				-40 & \frac{34}{3}     
                               \end{pmatrix}
\end{align}

Inserting these formulae into (\ref{equ:vecK}) we indeed verify that 
the coefficients $\vec{K}^\text{gauge}$ and $\vec{K}^\text{scalar}$ of 
logarithmic terms calculated by us are correct: they are consistent with 
the $\mu_{\rm in}$-independence of the resulting physical amplitudes at 
this order of perturbation theory.

It is instructive to compare the structure of the cancellation of the 
$\mu_{\rm in}$-dependence in the case of the gauge boson exchange with the 
one of the scalar exchange:
\begin{itemize}
\item
 In the gauge boson case the  $\mu_{\rm in}$-dependence of 
$\hat U(\mu_b,\,\mu_{\rm in})$ can only be cancelled by the corrections 
calculated by us. 
\item
The case of scalar exchange with LR couplings is quite different.
Here the $\mu_{\rm in}$-dependence of $\hat U(\mu_b,\,\mu_{\rm in})$ is totally 
cancelled by the one of the $ m^2(\mu_{\rm in})$ so that even without 
our corrections the amplitudes are $\mu_{\rm in}$ independent. Indeed 
the $C_{1,2}^{\rm LR}$ coefficients in the scalar case do not contain any 
logarithmic terms at $\ord(\alpha_s)$. 
This type of cancellation can be traced to the fact that the anomalous 
dimension of the $Q_2^{\rm LR}$ operator is as seen in Eq.~(\ref{LRQ}) 
up to the sign equal twice the 
anomalous dimension of the mass operator. The role of our calculation is 
then the removal of the renormalization scheme dependence.
\item
In the case of SLL operators the cancellation in question is not as pronounced 
because as seen in  Eq.~(\ref{SLLQ}) the anomalous dimensions of the relevant 
operators receive additional contributions beyond $\gamma^{(0)}_m$. In this 
case our calculation removes both scale and renormalization scheme dependences.
\end{itemize}


\boldmath
\section{Mixing Amplitudes at the NLO Level}\label{Sec5}
\unboldmath
\subsection{Preliminaries}
Having calculated  $\ord(\alpha_s)$ corrections to the Wilson coefficients 
at the matching scale  $\mu_{\rm in}$ we can obtain the complete NLO 
expressions for various tree-level contributions to the off-diagonal 
elements $M_{12}$ for the $K$ and $B_{s,d}$ systems. We will present only explicit expressions for $\Delta S=2$ transition. Analogous expressions for 
$B_{s,d}$ systems can be easily obtained in the same manner.

In presenting our results we will use
the so-called $P_i^a$ QCD factors of \cite{Buras:2001ra} that include both hadronic matrix
elements of contributing operators and renormalization group evolution from high energy to low energy scales. These factors depend on the system considered, 
depend on the high energy matching scale and depend on the renormalization 
scheme used to renormalize the operators. The formulae for these factors
have been given in \cite{Buras:2001ra} in the NDR-$\overline{\text{MS}}$ renormalization scheme 
of \cite{Buras:1989xd}. This scheme dependence is cancelled by the non-logarithmic $\ord(\alpha_s)$ corrections calculated by us. The 
logarithmic corrections cancel the scale dependence of $P^a_i$ as explained 
in the previous section. 

The formulae for various contributions to $M_{12}$ are easily obtained from 
the Hamiltonians presented in Section~\ref{Sec3} by replacing the operators by the 
corresponding  hadronic matrix elements in Eq.~(\ref{eq:matrix}) denoted there 
shortly by $\bar P^a_i(\mu_{\rm in})$. For completeness we recall the formulae 
for $P^a_i(\mu_{\rm in})$ \cite{Buras:2001ra}:
\be
\label{p1_factor}
P_1^{\rm VLL}(\mu_{\rm in})=\left[\eta (\mu_L,\mu_{\rm in})\right]_{\rm VLL} 
B^{\rm VLL}_1(\mu_L),
\ee
\be
P_1^{\rm LR}(\mu_{\rm in}) =-\frac{1}{2} \left[\eta_{11} (\mu_L,\mu_{\rm in})\right]_{\rm LR}
               \left[B^{\rm LR}_1(\mu_L)\right]_{\rm eff}
+\frac{3}{4} \left[\eta_{21} (\mu_L,\mu_{\rm in})\right]_{\rm LR}
               \left[B^{\rm LR}_2(\mu_L)\right]_{\rm eff},
\ee
\be
P_2^{\rm LR}(\mu_{\rm in})=-\frac{1}{2} \left[\eta_{12} (\mu_L,\mu_{\rm in})\right]_{\rm LR}
               \left[B^{\rm LR}_1(\mu_L)\right]_{\rm eff}
+\frac{3}{4} \left[\eta_{22} (\mu_L,\mu_{\rm in})\right]_{\rm LR}
               \left[B^{\rm LR}_2(\mu_L)\right]_{\rm eff},
\ee

\be
P_1^{\rm SLL}(\mu_{\rm in}) =-\frac{5}{8} \left[\eta_{11} (\mu_L,\mu_{\rm in})\right]_{\rm SLL}
               \left[B^{\rm SLL}_1(\mu_L)\right]_{\rm eff}
-\frac{3}{2} \left[\eta_{21} (\mu_L,\mu_{\rm in})\right]_{\rm SLL}
               \left[B^{\rm SLL}_2(\mu_L)\right]_{\rm eff},
\ee
\be
P_2^{\rm SLL}(\mu_{\rm in})=-\frac{5}{8} \left[\eta_{12} (\mu_L,\mu_{\rm in})\right]_{\rm SLL}
               \left[B^{\rm SLL}_1(\mu_L)\right]_{\rm eff}
-\frac{3}{2} \left[\eta_{22} (\mu_L,\mu_{\rm in})\right]_{\rm SLL}
               \left[B^{\rm SLL}_2(\mu_L)\right]_{\rm eff},
\ee
where $\mu_L$ is a low energy scale at which hadronic matrix elements 
are evaluated. Explicit formulae for the QCD-NLO factors 
$\left[\eta_{ij} (\mu_L,\mu_{\rm in})\right]_a$ are given in \cite{Buras:2001ra}.

The effective parameters
$\left[B^a_i(\mu_L)\right]_{\rm eff}$ are defined in the case of $K^0-\bar K^0$ 
mixing ($\mu_L=2\gev$) by 
\be
\label{Balat}
\left[B^a_i(\mu_L)\right]_{\rm eff}\equiv
\left(\frac{m_K}{m_s(\mu_L)+m_d(\mu_L)}\right)^2 B^a_i(\mu_L)=
25.61\left[\frac{98~{\rm MeV}}{m_s(\mu_L)+m_d(\mu_L)}\right]^2
B^a_i(\mu_L).
\ee
In the case of $B_{d,s}^0-\bar B_{d,s}^0$ mixings one has to make the replacements
$\mu_L\to \mu_b$ and $m^2_K F_K^2\to m^2_{B_q} F_{B_q}^2$. Then in the case of
$B_d^0-\bar B^0_d$ system (at $\mu_b = 4.6$) 
\be
\left[B^a_i(\mu_b)\right]_{\rm eff}\equiv
\left(\frac{m_B}{m_b(\mu_b)+m_d(\mu_b)}\right)^2 B^a_i(\mu_b)=
1.68\left[\frac{4.08~{\rm GeV}}{m_b(\mu_b)+m_d(\mu_b)}\right]^2
B^a_i(\mu_b),
\ee
with an analogous formula for the $B_s^0-\bar B^0_s$ system.

We list now the final
NLO expressions for the mixing amplitudes.
\subsection{Colourless gauge boson}

 \begin{align}\begin{split}
  M_{12}^\star(\Delta S = 2)  =&
 \frac{(\Delta_L^{sd}(A))^2}{2M_A^2}C_1^\text{VLL}(\mu_A)\bar P_1^\text{VLL}(\mu_A)
 +\frac{(\Delta_R^{sd}(A))^2}{2M_A^2}
 C_1^\text{VRR}(\mu_A)\bar P_1^\text{VLL}(\mu_A) \\
 &+\frac{\Delta_L^{sd}(A)\Delta_R^{sd}(A)}{
 M_A^2} \left [ C_1^\text{LR}(\mu_A) \bar P_1^\text{LR}(\mu_A) +
 C_2^\text{LR}(\mu_A) \bar P_2^\text{LR}(\mu_A) \right]\,.\end{split}
 \end{align}

The relevant Wilson coefficients are given in Section~\ref{CLGB}.

\subsection{Colourless scalar}

 \begin{align}\begin{split}
  M_{12}^\star(\Delta S = 2) =&
 -\frac{(\Delta_L^{sd}(H))^2}{2M_H^2}\left[C_1^\text{SLL}(\mu_H)
 \bar P_1^\text{SLL}(\mu_H) +C_2^\text{SLL}(\mu_H)\bar P_2^\text{SLL}(\mu_H)
 \right]\\
 &-\frac{(\Delta_R^{sd} (H))^2 } {2 M_H^2 }
 \left[C_1^\text{SRR}(\mu_H)\bar P_1^\text{SRR}(\mu_H) 
 +C_2^\text{SRR}(\mu_H)\bar P_2^\text{SRR}(\mu_H) \right]\\
 &-\frac{\Delta_L^{sd}(H)\Delta_R^{sd}(H)}{
 M_H^2} \left [ C_1^\text{LR}(\mu_H) \bar P_1^\text{LR}(\mu_H) +
 C_2^\text{LR}(\mu_H) \bar P_2^\text{LR}(\mu_H) \right]\,.\end{split}
 \end{align}
The relevant Wilson coefficients are given in Section~\ref{CLS}.

We would like to emphasize that the formulae of this Section together with
the QCD factors $\eta_{ij}$ presented in Section 3 of \cite{Buras:2001ra} 
and the coefficients  $C^a_i(\mu_{\rm in})$ calculated in Section~\ref{Sec3} 
of the present paper
 are valid for the tree level contributions of colourless bosons and 
scalars in
{\it any} extension of the SM in which such contributions are present. The 
only model dependence enters through the couplings $\Delta_{L,R}^{ij}$ 
and the gauge boson and scalar masses.
In particular the coefficients 
$P^a_i$ are universal in a given meson system and 
given renormalization scheme. With our normalization of $C^a_i(\mu_{\rm in})$ also 
these coefficients are universal except for the scale $\mu_{\rm in}$ of NP 
and the same renormalization scheme used to evaluate $P^a_i$.
In the process of multiplying
 $C^a_i(\mu_{\rm in})$ and $P^a_i$ terms ${\cal O}(\alpha_s^2)$ have to
be removed.

\section{Numerical Analysis}\label{Sec6}

We will now compute the size of $\ord(\alpha_s)$ corrections and their impact
on the reduction of the unphysical $\mu_{\rm in}$-dependence present in 
the analyses  in the literature. It should be 
recalled that the actual size of the corrections calculated by us is not 
the most important result as these corrections are given in a particular 
renormalization scheme, the NDR-$\overline{\text{MS}}$ scheme. They could 
be different in another scheme. But then also the $P_i^a$ factors would 
be different so that the final result for the physical mixing amplitudes 
$M_{12}$ would be renormalization scheme independent up to $\ord(\alpha_s^2)$ 
corrections, that is NNLO corrections. Thus the important result of our 
paper is that we provide for the first time  
mixing amplitudes resulting from 
tree level decays including NLO QCD corrections 
that are renormalization scheme independent and 
which do not depend on a precise choice of the matching scale. Both 
statements are valid up to NNLO corrections.

There are four linear combinations of $P_i^a$ and of the $C_i^a$ that 
should be scale and renormalization scheme independent. As we 
normalized the non-vanishing $C_i^a$ at LO to unity, these combinations 
are model independent. The model dependence enters only through the 
fermion-boson
couplings and heavy boson masses characteristic for a given model.
The linear combinations in question in the case of the $K$ system 
 are given as follows:

{\bf Gauge Bosons:}

\begin{align}
 R_1^K &=C_1^\text{VLL}(\mu_A) P_1^\text{VLL}(\mu_A)\,,\label{R1}\\
R_2^K&=C_1^\text{LR}(\mu_A)  P_1^\text{LR}(\mu_A) +
 C_2^\text{LR}(\mu_A)  P_2^\text{LR}(\mu_A)\, ,\label{R2}
\end{align}
with the $P_i^a$ factors evaluated using hadronic $K^0-\bar K^0$ 
matrix elements at scale $\mu_L=2\gev$.

{\bf Scalars:}

\begin{align}
 R_3^X & = \left(C_1^\text{SLL}(\mu_H)
  P_1^\text{SLL}(\mu_H) +C_2^\text{SLL}(\mu_H) P_2^\text{SLL}(\mu_H)\right)\frac{m^2(\mu_H)}{m^2(M_H)}\,,\label{R3}\\
R_4^X & =\left(C_1^\text{LR}(\mu_H) P_1^\text{LR}(\mu_H) +
 C_2^\text{LR}(\mu_H)  P_2^\text{LR}(\mu_H)\right)\frac{m^2(\mu_H)}{m^2(M_H)} \,.\label{R4}
\end{align}

Note that even if the quantities $R_2$ and $R_4$ appear at first sight 
to be the same they differ from each other because the Wilson coefficients of 
the $Q_{1,2}^{\rm LR}$ operators
for the gauge boson case are different than for the Higgs case.

\begin{figure}[!tbh]
\centering

\includegraphics[width=.48\linewidth]{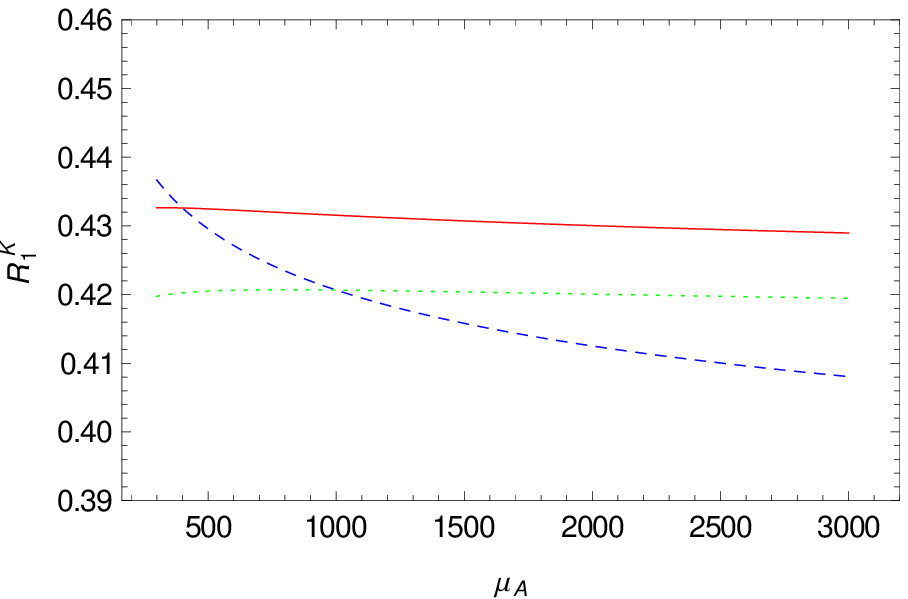}
\hspace{0.2cm}
\includegraphics[width=.48\linewidth]{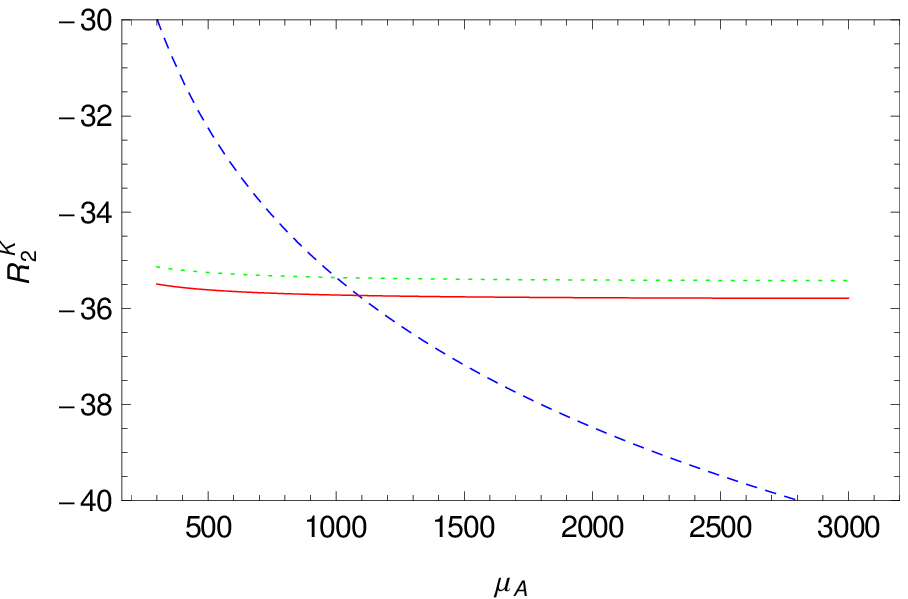}

\includegraphics[width=.48\linewidth]{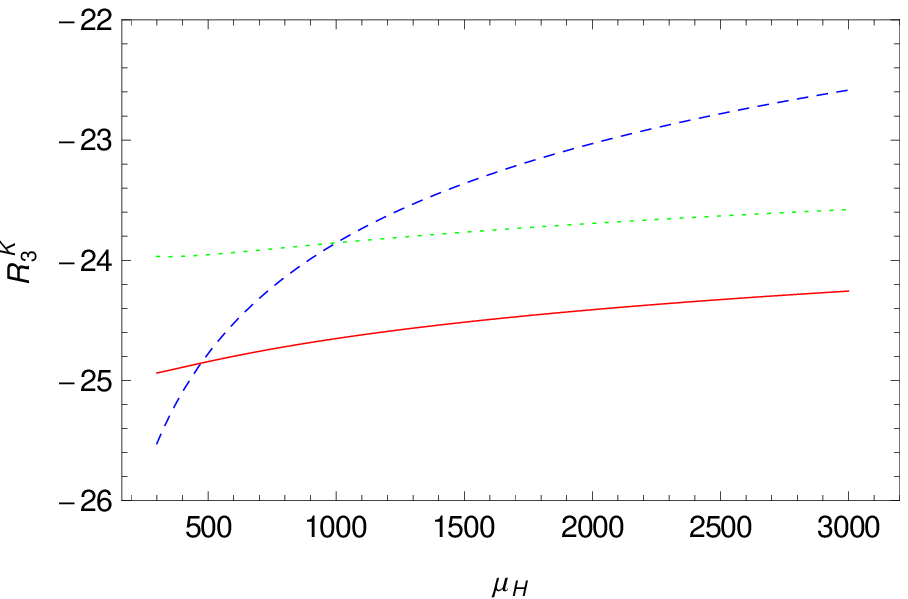}
\hspace{0.2cm}
\includegraphics[width=.48\linewidth]{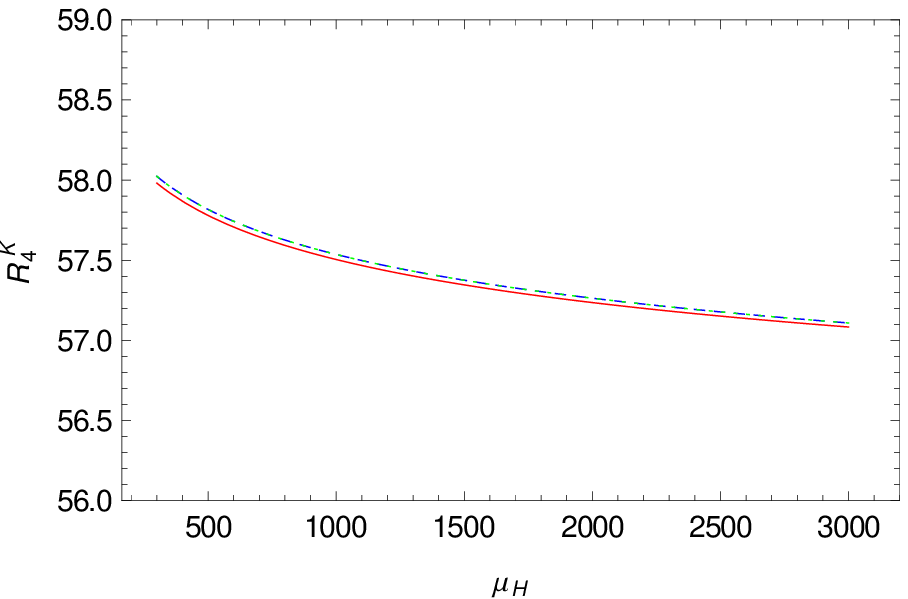}
\caption{ \it The quantities $R_i^K$ defined in Eq.~(\ref{R1})--(\ref{R4}) as a function of $\mu_{A,H}$ for $M_{A,H} = 1~$TeV. In the
first row are the
results for the gauge boson exchange  and in the second row for scalar
exchange. The dashed blue line represents the LO calculation for the Wilson coefficients, the dotted green line includes only logarithmic
$\mathcal{O}(\alpha_s)$ corrections which reduces the scale dependence and the solid red line includes both logarithmic
and non-logarithmic $\mathcal{O}(\alpha_s)$ corrections  that is important for the cancellation of scale and
scheme
dependence.}\label{fig:plots}
\end{figure}

In Figs.~\ref{fig:plots} we plot $R_i^K$  as functions of the matching scales setting as 
an example the masses of gauge bosons and scalars to $1\tev$. Since the  
$\mu_{\rm in}$ dependence is the same in the $B$ and $K$ system we do
not
show the results for  $B$ mesons. They only differ in magnitude from each other
because the hadronic matrix elements hidden in $P_i^a$ are different in these two 
meson sectors. Moreover, QCD effects are generally stronger in the $K$ 
system because the renormalization group evolution is over a larger 
range of scales.

In order to find numerical values of $P_i^a$ one needs the values of 
the corresponding non-perturbative parameters $B^a_i$ defined in  
\cite{Buras:2001ra}. These are given in terms of the parameters $B_i$
used in \cite{Ciuchini:1997bw,Ciuchini:1998ix,Babich:2006bh,Becirevic:2001xt} as follows:
\begin{subequations}
\begin{align}
& B_1^\text{VLL}(\mu_0) = B_1^\text{VRR}(\mu_0) = B_1(\mu_0)\,,\\
& B_1^\text{LR}(\mu_0) = B_5(\mu_0)\,,\\
& B_2^\text{LR}(\mu_0) = B_4(\mu_0)\,,\\
& B_1^\text{SLL}(\mu_0) = B_2^\text{SRR}(\mu_0) = B_2(\mu_0)\,,\\
& B_2^\text{SLL}(\mu_0) = B_2^\text{SRR}(\mu_0) = \frac{5}{3}B_2(\mu_0) - \frac{2}{3}B_3(\mu_0)\,.
\end{align}
\end{subequations}

The values for $B_i$ in the $\overline{\rm MS}$-NDR scheme extracted from~\cite{Babich:2006bh} 
for the $K^0-\bar K^0$ system are collected in Table~\ref{tab:B_i}, together with the relevant value of $\mu_0$. 

\begin{table}[!ht]
{\renewcommand{\arraystretch}{1.3}
\begin{center}
\begin{tabular}{|c|c|c|c|c|c|c|}
\hline
&$B_1$&$B_2$& $B_3$&$B_4$&$B_5$&$\mu_0$\\
\hline
$K^0$-$\bar K^0$&0.57&0.68&1.10&0.81&0.56&2.0\gev\\
\hline
\end{tabular}
\end{center}}
\caption{\it Values of the parameters $B_i$ in the $\overline{\text{MS}}$-NDR scheme obtained in~\cite{Babich:2006bh} for $K^0-\bar K^0$ system. The scale $\mu_0$ at which $B_i$ are evaluated is given in the last column. 
\label{tab:B_i}}
\end{table}

In each case we vary conservatively the matching scale between $300\gev$ 
and $3000\gev$ and show in each plot three curves:
\begin{itemize}
\item
The result without the inclusion of $\ord(\alpha_s)$ corrections as used 
until now in the literature (blue, dashed line).
\item
The result after only logarithmic terms in the $\ord(\alpha_s)$ have 
been included. They are crucial for the cancellation of the matching scale
dependence (green, dotted line).
\item
The result after the inclusion of non-logarithmic $\ord(\alpha_s)$ 
corrections that are crucial
for the cancellation of the renormalization scheme dependence (red, solid line).
\end{itemize}

These plots are self-explanatory and we make only a few comments:
\begin{itemize}
\item 
As expected the dotted green lines cross the dashed blue lines at 
$\mu_{A,H}=1000\gev$.
\item
The crossing point between the  solid red  and dashed blue  lines is generally 
at $\mu_{A,H}$ that differs from the mass $M_{A,H}=1000\gev$ of the exchanged 
particle. The modest size of these differences shows that these 
corrections are small, of $\ord(5)\%$ in 
the  NDR-$\overline{\text{MS}}$ scheme.
\item
In the case of gauge boson exchanges the matching scale dependence 
of roughly $10 (25)\%$ for VLL(LR) operators in the range considered,  when
$\ord(\alpha_s)$ corrections are not taken into account,  has been reduced down to 
$1-2\%$ after the inclusion of these corrections.
\item
In the case of scalar exchanges the matching scale dependence in the case 
of the SLL sector
of roughly $10\%$ in the range considered, when
$\ord(\alpha_s)$ corrections  are not taken into account,
has been reduced down to 
$1-2\%$ after the inclusion of $\ord(\alpha_s)$ corrections.
\item
In the case of LR operators in the scalar case there is basically no 
left-over scale dependence even in the LO as we explained at the end of 
Section~\ref{Sec4}.
\end{itemize}

This reduction of scale uncertainties cannot be appreciated at present 
in view of significant uncertainties in the values of the $B_i$ 
parameters, as seen in Table~\ref{tab:B_i}, but the recent advances in 
lattice calculations allow for optimism and 
we expect that during this decade these uncertainties 
could be reduced below $5\%$ and then the calculations presented here 
will turn out to be important.

\section{Summary}\label{Sec7}
If there is a new physics at  distance scales as short as $10^{-19}-10^{-21}$~m, 
it will manifest itself primarily not through penguin and box diagrams as in 
the SM but through  tree level FCNC processes. 
The best known examples of such NP are
various versions of the so-called $Z^\prime$ models
in which new neutral heavy weak boson ($Z^\prime$) 
mediate FCNC processes already at tree level. 
Gauged flavour models with new very heavy neutral 
gauge bosons 
and Left-Right symmetric models with 
heavy neutral scalars are other prominent examples where tree-level 
contributions to
$\Delta F=2$ amplitudes are present.

Effective tree-level contributions to $\Delta F=2$ observables can  also be generated at one loop in models 
having GIM at the fundamental level and Minimal Flavour Violation of which 
Two-Higgs Doublet Models with and without supersymmetry are the best known 
examples. In models with heavy vectorial fermions that mix with the standard 
chiral quarks 
and models in which $Z^0$ and SM neutral Higgs $H^0$ mix with new heavy 
gauge bosons and scalars in the process of electroweak symmetry breaking also 
tree-level  contributions to $\Delta F=2$  processes mediated by $Z^0$ and SM neutral Higgs $H^0$ are possible. In all these extensions new local operators absent in 
the SM are generated having Wilson coefficients that are generally much 
stronger affected by renormalization group QCD effects than it is the case 
of the SM operators.

Present studies of renormalization group QCD effects performed at the NLO 
level in many extensions of the SM use  the so-called $P_i^a$ QCD factors 
\cite{Buras:2001ra} that include both hadronic matrix
elements of contributing operators and renormalization group evolution from high energy to low
energy scales. These factors represent the dominant part of any  NLO QCD analysis 
but do not take into account $\ord(\alpha_s)$ corrections to Wilson coefficients at the matching scale which separates the full and effective theories. 
Therefore basically all published calculations that considered tree level 
decays suffer from some unphysical scale and renormalization scheme
dependences. While presently these unphysical effects are much smaller than 
the uncertainties in the hadronic matrix elements of contributing operators, 
the situation could change in this decade due to important progress in lattice 
simulations with dynamical fermions \cite{Antonio:2007pb,Aubin:2009jh,Laiho:2009eu,Bae:2010ki,Constantinou:2010qv,Aoki:2010pe,McNeile:2011ng,Bouchard:2011xj,Neil:2011ku}.

While a general calculations of $\ord(\alpha_s)$ corrections to Wilson 
coefficients,  when the leading contributions 
come from loop diagrams, is very model dependent, 
a rather general analysis can be done for 
tree level exchanges so that the final results depend only on the 
couplings of exchanged bosons (vectors and scalars), on the QCD colour 
factors and the QCD coupling constant.

The main goal of our analysis was to provide analytical formulae for 
arbitrary number of colours ($N$) for the $\ord(\alpha_s)$ corrections 
in question in the case of
tree level $\Delta F=2$ processes mediated by heavy colourless gauge bosons and scalars. The results for the Wilson Coefficients 
and effective Hamiltonians for these cases are 
collected in Section~\ref{Sec3}, while the corresponding mixing amplitudes 
at the NLO level that combine our results with the known $P_i^a$ QCD factors 
are presented in Section~\ref{Sec5}. In Section~\ref{Sec6} we demonstrated 
that the unphysical scale dependences have practically been removed by 
our calculations. This is particularly important for the LR system in the 
case of gauge boson exchanges, where the scale dependence at LO is sizeable. 
The Appendices collect certain technicalities about 
the evanescent operators and intermediate results in 
the full and effective theories which should enable interested readers to
check our calculations.

\subsection*{Acknowledgements}
We would like to thank Gerhard Buchalla, Andreas Kronfeld, Mikolaj Misiak and Ulrich 
Nierste for discussions. 
This research was done in the context of the ERC Advanced Grant project ``FLAVOUR''(267104) and was partially supported by the DFG cluster
of excellence 
``Origin and Structure of the Universe''.


\appendix

\section{The issue of Evanescent Operators}
It is well known that in the process of NLO calculations in the NDR-$\overline{\text{MS}}$ scheme, where ultraviolet divergences are regulated dimensionally,
the so-called 
evanescent operators that vanish in $D=4$ dimensions have to be considered. 
They arise in particular when complicated Dirac structures are projected 
onto the chosen basis of physical operators.
The treatment of this operators in the process of matching considered by us 
must be consistent with the one used in the calculation of two-loop 
anomalous dimensions. 

We have used the $P_i^a$ QCD factors from \cite{Buras:2001ra} which were
based on the two-loop anomalous dimensions of operators calculated 
in \cite{Buras:2000if}. Therefore 
it is mandatory for us to treat evanescent operators  appearing in our 
calculations
in the same manner 
as done in  \cite{Buras:2000if}. Now, the latter paper used the treatment 
of evanescent operators as proposed in the context of the formulation of the 
 NDR-$\overline{\text{MS}}$ scheme introduced in \cite{Buras:1989xd}. The virtue of 
this treatment is that the evanescent operators defined in this scheme 
influence only two-loop anomalous dimensions. By definition they do not 
contribute to the matching and to the finite corrections to the matrix elements  of renormalized operators calculated by us. They are simply subtracted away in
the process of renormalization. This issue is summarized in Section 6.9.4 of 
\cite{Buras:1998raa}, where further references can be found. 
A very important paper in this context is \cite{Herrlich:1994kh}. Therefore effectively the  
calculations presented here were based on the projections listed in the next appendix that 
leave out the evanescent operators on the r.h.s. 

 In this context we should warn the reader that the  NDR-$\overline{\text{MS}}$ scheme used in \cite{Beneke:1998sy}, while sharing
all the virtues of the scheme of \cite{Buras:1989xd} uses different projections in the SLL sector. This implies 
 different 
two-loop anomalous dimensions for the SLL operators, that is also different 
$P_i^{\rm SLL}$ and also different $\ord(\alpha_s)$ terms in $C_i^{\rm SLL}(\mu_{\rm in})$ so that the physical amplitudes are the same in both schemes. The relation between 
these two schemes has been worked out in \cite{Gorbahn:2009pp}.

Another technical issue is related to the Fierz-vanishing evanescent operators, 
which have to be considered 
when one wants to relate the operators with non-singlet structures like 
\bea
\tilde Q_1^{\rm SLL} &=& (\bar s^\alpha P_L d^\beta )
                     (\bar s^\beta  P_L d^\alpha),\\
\tilde Q_2^{\rm SLL} &=& (\bar s^\alpha \sigma_{\mu\nu} P_L d^\beta )
                     (\bar s^\beta  \sigma^{\mu\nu} P_L d^\alpha)
\eea
to the operators $Q_{1,2}^{\rm SLL}$ used by us. In $D=4-2\epsilon$ dimensions 
the usual $D=4$ identities
\be \label{opfSLL1}
\tilde Q_1^{\rm SLL} \bbuildrel{=}_{D=4}^{} -\frac{1}{2}\,Q_1^{\rm SLL}
+\frac{1}{8}\,Q_2^{\rm SLL},
\ee
\be\label{opfSLL2}
\tilde Q_2^{\rm SLL} \bbuildrel{=}_{D=4}^{} 6\,Q_1^{\rm SLL}+\frac{1}{2}\,Q_2^{\rm SLL}. 
\ee
do not work and one has to add evanescent operators on the r.h.s of these 
equations. However again, even if the inclusion of the latter operators was 
relevant for the calculation of two-loop anomalous dimensions of the 
physical operators in 
\cite{Buras:2000if}, it turns out that they do not contribute to the matching 
as long as the infrared divergences are not regulated dimensionally. As in 
our paper we regulated such divergences by a non-vanishing $p^2$, we can use 
the relations in Eq.~(\ref{opfSLL1}) and Eq.~(\ref{opfSLL2}) without taking the Fierz-vanishing 
evanescent operators 
in question into account. As discussed in \cite{Buras:2000if} these 
``problems'' are absent in the case of other operators.

\section{Projections}
We list projections of all Dirac structures on physical operators that 
we encountered in our calculations. These projections correspond to the 
so-called ``Greek method'' as described in Section 6.9 of  \cite{Buras:1989xd}. 
The evanescent operators, relevant in this renormalization scheme only 
at two-loop level, are defined as the differences of l.h.s and r.h.s of 
these equations. See \cite{Buras:2000if} for more details. We define $\sigma_{\mu\nu} = \frac{1}{2}\left[\gamma_\nu,\,\gamma_\nu\right]$.

\begin{subequations}
\begin{align}
&\gamma_\alpha\gamma_\beta \gamma_\mu (1\pm \gamma_5) \gamma^\beta\gamma^\alpha  \otimes  \gamma^\mu (1\pm
\gamma_5)  = 4 (1-2\epsilon)\,\gamma_\mu
(1\pm \gamma_5)
\otimes  \gamma^\mu (1\pm \gamma_5)\\
& \gamma_\mu (1\pm \gamma_5) \gamma_\alpha\gamma_\beta \otimes \gamma^\mu (1\pm
\gamma_5)\gamma^\alpha\gamma^\beta =      4 (4-\epsilon)\,\gamma_\mu (1\pm \gamma_5) \otimes\gamma^\mu (1\pm
\gamma_5)\\
&\gamma_\mu (1\pm \gamma_5) \gamma_\alpha\gamma_\beta \otimes \gamma^\beta\gamma^\alpha  \gamma^\mu (1\pm
\gamma_5)  = 4 (1-2\epsilon)\,\gamma_\mu (1\pm \gamma_5) \otimes  \gamma^\mu (1\pm \gamma_5)
\end{align}
\end{subequations}

\begin{subequations}
\begin{align}
&\gamma_\alpha\gamma_\beta \gamma_\mu (1\pm \gamma_5) \gamma^\beta\gamma^\alpha  \otimes  \gamma^\mu (1\mp
\gamma_5)   
  = 4 (1-2\epsilon)\,\gamma_\mu (1\pm \gamma_5)
\otimes \gamma^\mu (1\mp \gamma_5)\\
& \gamma_\mu (1\pm \gamma_5) \gamma_\alpha\gamma_\beta \otimes \gamma^\mu (1\mp \gamma_5)
\gamma^\alpha\gamma^\beta =     4 (1+\epsilon)\,\gamma_\mu (1\pm \gamma_5) \otimes \gamma^\mu (1\mp
\gamma_5)\\
&\gamma_\mu (1\pm \gamma_5) \gamma_\alpha\gamma_\beta \otimes \gamma^\beta\gamma^\alpha  \gamma^\mu (1\mp
\gamma_5)=  \,  16 (1-\epsilon)\gamma_\mu (1\pm \gamma_5) \otimes \gamma_\mu (1\mp \gamma_5)
\end{align}
\end{subequations}

\begin{subequations}
\begin{align}
& \gamma_\nu\gamma_\mu(1\mp\gamma_5)\gamma^\mu\gamma^\nu\otimes(1\pm\gamma_5) = 16
(1-\epsilon)\,(1\mp\gamma_5)\otimes (1\pm\gamma_5)\\
& (1\mp\gamma_5)\gamma_\mu\gamma_\nu\otimes(1\pm\gamma_5)\gamma^\mu\gamma^\nu = 4
(1+\epsilon)\,(1\mp\gamma_5)\otimes (1\pm\gamma_5)\\
& (1\mp\gamma_5)\gamma_\mu\gamma_\nu\otimes\gamma^\mu\gamma^\nu(1\pm\gamma_5) = 4
(1-2\epsilon)\,(1\mp\gamma_5)\otimes (1\pm\gamma_5)
\end{align}
\end{subequations}

\begin{subequations}
\begin{align}
 &\gamma_\nu\gamma_\mu(1\pm\gamma_5)\gamma^\mu\gamma^\nu\otimes(1\pm\gamma_5) = 16(1-\epsilon)\,
(1\pm\gamma_5)\otimes (1\pm\gamma_5)\\
\begin{split}
 &(1\pm\gamma_5)\gamma_\mu\gamma_\nu\otimes(1\pm\gamma_5)\gamma^\mu\gamma^\nu
\\
&\quad
= (4-2\epsilon)\,
(1\pm\gamma_5)\otimes (1\pm\gamma_5) +\sigma_{\mu\nu}(1\pm\gamma_5)\otimes
\sigma^{\mu\nu}(1\pm\gamma_5)\end{split}\\
\begin{split}
& (1\pm\gamma_5)\gamma_\mu\gamma_\nu\otimes\gamma^\nu\gamma^\mu(1\pm\gamma_5)\\
&\quad= (4-2\epsilon)\,
(1\pm\gamma_5)\otimes (1\pm\gamma_5)-\sigma_{\mu\nu}(1\pm\gamma_5)\otimes
\sigma^{\mu\nu}(1\pm\gamma_5)\end{split}
\end{align}
\end{subequations}

\begin{subequations}
\begin{align}
&  \gamma^\alpha\gamma^\beta\sigma^{\mu\nu}(1\pm\gamma_5)\gamma_\beta\gamma_\alpha \otimes
\sigma^{\mu\nu}(1\pm\gamma_5) = 0\\
\begin{split}
 &\sigma_{\mu\nu}(1\pm\gamma_5) \gamma_\alpha\gamma_\beta \otimes \sigma^{\mu\nu}(1\pm\gamma_5)
\gamma^\alpha \gamma^\beta  \\
&\quad=(48-80\epsilon) \,(1\pm\gamma_5)\otimes (1\pm\gamma_5) +(12-6\epsilon)\,
\sigma_{\mu\nu}(1\pm\gamma_5)\otimes \sigma^{\mu\nu}(1\pm\gamma_5)\end{split}\\
\begin{split}
 &\sigma_{\mu\nu}(1\pm\gamma_5) \gamma_\alpha\gamma_\beta \otimes\gamma^\beta\gamma^\alpha
\sigma^{\mu\nu}(1\pm\gamma_5)  \\
&\quad=-(48-80\epsilon)\, (1\pm\gamma_5)\otimes
(1\pm\gamma_5) +(12-14\epsilon)\,
\sigma_{\mu\nu}(1\pm\gamma_5)\otimes \sigma^{\mu\nu}(1\pm\gamma_5)\end{split}
\end{align}
\end{subequations}

\section{Matrix Elements of Operators}

After quark wave function renormalization and operator renormalization we get
{\allowdisplaybreaks
\begin{align}
 \langle Q_1^\text{VLL}\rangle &= \left[ 1 + 2 C_F \frac{\alpha_s}{4\pi}
\log\frac{\mu^2}{-p^2}-3\frac{N-1}{N}\frac{\alpha_s}{4
\pi} \left( \log\frac{\mu^2}{-p^2}+\frac{7}{3}\right)\right] Q_1^\text{VLL}\\
\begin{split}
 \langle Q_1^\text{LR}\rangle &=\left[1+2 C_F \frac{\alpha_s}{4\pi}
\log\frac{\mu^2}{-p^2} \right]Q_1^\text{LR}   -6\frac{\alpha_s}{4\pi}\left(
\log\frac{\mu^2}{-p^2} + \frac{1}{3}\right)\left(\frac{1}{2N}Q_1^\text{LR}+ Q_2^\text{LR}\right) 
\end{split}\\
\begin{split}
 \langle Q_2^\text{LR}\rangle &=\left[1+8 C_F \frac{\alpha_s}{4\pi}\left(1+
\log\frac{\mu^2}{-p^2}\right) \right]Q_2^\text{LR}  +6\frac{\alpha_s}{4\pi}\left(\frac{1}{2N}Q_2^\text{LR}+
\frac{1}{4}Q_1^\text{LR}\right) 
\end{split}\\
\begin{split}
  \langle Q_1^\text{SLL}\rangle &=\left[1+8 C_F \frac{\alpha_s}{4\pi}\left(1+
\log\frac{\mu^2}{-p^2}\right) \right]Q_1^\text{SLL}\\
&\quad - \frac{\alpha_s}{4\pi}\left(2+
\log\frac{\mu^2}{-p^2}\right)\left(\frac{N-2}{4N} Q_2^\text{SLL}+ 3 Q_1^\text{SLL}\right)\end{split}\\
\begin{split}
  \langle Q_2^\text{SLL}\rangle &= Q_2^\text{SLL} - 48 \frac{\alpha_s}{4\pi}\left(
\log\frac{\mu^2}{-p^2}+\frac{1}{3}\right)\left(-\frac{N+2}{4N} Q_1^\text{SLL}+\frac{1}{16}
Q_2^\text{SLL}\right)\\
&\quad -4 \frac{\alpha_s}{4\pi}\left(\frac{N-2}{4N} Q_2^\text{SLL}+ 3 Q_1^\text{SLL}\right)\end{split}
\end{align}}%

We remark that for the matching performed in this paper only 
the $\ord(\alpha_s)$ corrections to the matrix 
elements $\langle Q_1^\text{VLL}\rangle$, $\langle Q_1^\text{LR}\rangle$ 
and $\langle Q_1^\text{SLL}\rangle$ matter. We give the remaining matrix 
elements as they are relevant for NLO matching when coloured gauge bosons 
and scalars are exchanged \cite{BG2}.

\section{Amplitudes in the Full Theory}

Colourless gauge boson exchange (after quark wave function renormalization):
\begin{align}
 (\mathcal{A}^\text{VLL}_1)^\text{gauge} = & \frac{(\Delta_L^{sd}(A))^2}{2 M_A^2}\left[1+ 2
C_F\frac{\alpha_s}{4\pi}
\log\frac{\mu^2}{-p^2}- 3 \frac{N-1}{N} \frac{\alpha_s}{4\pi}\left(\frac{1}{2}+ \log\frac{M^2_A}{-p^2}\right) \right]
Q_1^\text{VLL}\\
\begin{split}
 (\mathcal{A}^\text{LR}_1)^\text{gauge} =&  \frac{\Delta_L^{sd}(A)\Delta_R^{sd}(A)}{M_A^2}\left[\left(1+ 2
C_F\frac{\alpha_s}{4\pi}
\log\frac{\mu^2}{-p^2} \right)Q_1^\text{LR}\right.\\
& \qquad\qquad\qquad\quad\left. - \frac{\alpha_s}{4\pi}\left(\frac{1}{2}+
\log\frac{M^2_A}{-p^2}\right)\left(\frac{3}{N}Q_1^\text{LR}+6 Q_2^\text{LR}\right) \right]
\end{split}
\end{align}

Colourless scalar exchange (after quark wave function and quark mass renormalization):
\begin{align}
 (\mathcal{A}^\text{LR}_2)^\text{scalar} = & -\frac{\Delta_L^{sd}(H)\Delta_R^{sd}(H)}{M_H^2}\left[1 +
8 C_F \frac{\alpha_s}{4\pi} \left( 1+ \log\frac{\mu^2}{-p^2}\right)  \right]Q_2^\text{LR}\\
\begin{split}
(\mathcal{A}^\text{SLL}_1)^\text{scalar}  = &-\frac{(\Delta_L^{sd}(H))^2}{2 M_H^2}\left[ \left(1 + 8
C_F\frac{\alpha_s}{4\pi}\left(1+
\log\frac{\mu^2}{-p^2}\right) \right)Q_1^\text{SLL}\right.\\
&\qquad\qquad\qquad\quad\left.- \frac{\alpha_s}{4\pi} \left(\frac{1}{2}+\log\frac{M^2_H}{-p^2}\right)\left(\frac{N-2}{4N}Q_2^\text{SLL} + 3
Q_1^\text{SLL} \right) \right]
\end{split}
\end{align}

\bibliographystyle{JHEP}
\bibliography{LitBG}

\end{document}